\documentstyle[prl,aps]{revtex}
\begin{document}

\centerline{\large \bf Non-Dissipative Logic Device NOT}
\centerline{\large \bf Based on Two Coupled Quantum Dots}

\vskip 2mm

\centerline{L.A.Openov and A.M.Bychkov}

\vskip 2mm

\centerline{\it Moscow State Engineering Physics Institute
(Technical University),}
\centerline{\it Moscow, 115409, Russia}

\vskip 4mm

\begin{quotation}

Non-dissipative dynamics of interacting electrons in two
tunnel-coupled quantum dots is studied theoretically within the framework of
the Hubbard model. Various values of intra-dot Coulomb repulsion energy $U$
and inter-dot tunneling energy $V$ are considered, which correspond to
various size of the dots and to various distance between them. In the ground
state, the average value of the spin projection (magnetic moment) at each dot
is zero. The input signal (the local external magnetic field $H$) applied to
one of the dots at a time $t=0$ causes the electronic subsystem to evolve in
such a way that magnetic moments of quantum dots become oriented in the
opposite directions at any time $t>0$. For any set of $U$ and $V$, there
exist optimal values of $H$ and $t$ which maximize the absolute values of
magnetic moments at both dots, and magnetic moments become almost saturated.
Thus, the antiferromagnetic-like spin ordering can be realized at the stage
of coherent temporal evolution, well before the relaxation to a new ground
state at the sacrifice of inelastic processes. This effect ("dynamical
antiferromagnetism") may be used for implementation of a logic function NOT
in an extremely short time. A possibility to use the arrays of quantum dots
as high-speed single-electron devices of new generation is discussed.

\end{quotation}

\vskip 4mm

\centerline{\bf 1. Introduction}

\vskip 2mm

Feynman's original idea \cite{Feynman} to use the states of a quantum
mechanical system for information coding and processing has stimulated the
work on a quantum theory of computation \cite{Bennett}. This
theory takes advantage of fundamental properties of quantum systems, such as
superposition, interference, entanglement, nonlocality, and uncertainty.
The quantum bit ("qubit") can exist in arbitrary superpositions of classical
bits, 0 and 1. In this way it appears to be possible to develop very
efficient computational algorithms which are exponentially faster than
conventional ones \cite{Shor}. The first experimental demonstrations of
quantum gates (elementary units capable to perform simplest logical
operations) have been reported recently \cite{Turchette,Monroe}. They are
based on encoding the qubits in photon states. However, a lot of skepticism
has been given to a possibility of practical utilization of such gates in a
quantum computer \cite{Haroche,Landauer}.

Another way is to use quantum mechanical systems to employ {\it classical}
Boolean logic. For example, Lent {\it et al.} \cite{Lent} have proposed to
represent the bits 0 and 1 by two opposite charge polarizations of a cell
containing five quantum dots occupied by two electrons. Proper lateral
arrangements of such cells with respect to each other allow the realization
of various logical functions (NOT, AND, OR, {\it etc}). Nomoto {\it et
al.} \cite{Nomoto} have studied a logic device AND/OR composed of a pair of
tunnel-coupled quantum dots. They viewed unoccupation/occupation of a
quantum dot by a single electron as a bit 0/1. Bandyopadhyay {\it et al.}
\cite{Bandyo} have discussed a concept of {\it spin gates}. In that approach
the bits of information are carried by spins of individual electrons, e.g.,
the logical 1 (0) corresponds to the spin "up" ("down") direction of
electron spin at a given quantum dot. In arrays of quantum dots, the quantum
tunneling of electrons between adjacent dots and/or Coulomb interaction of
electrons with each other play a role of "wiring", resulting in the signal
propagation from dot to dot ('quantum coupled architecture').

Of course, the classical computational algorithms are not as efficient as
the quantum ones since they do not benefit from purely quantum effects
(superposition, interference, {\it etc.}). Nevertheless, one would expect
that the reduction in size of computer components will result in increase
of the computational speed (up to inverse atomic times) and data storage
density (up to several units per nm$^2$ for planar structures). Besides, the
Boolean logic seems to be more closer to a practical implementation at a
nanometer scale. Indeed, a number of techniques have been developed to
produce arrays of quantum dots on a solid surface (see, e.g., \cite{Bandyo,%
Mamin,Meurer}). In addition, the scanning tunneling microscopy provides a
direct way to write-in and read-out the information since the tunnel current
is sensitive to the magnetic moment of a single atom \cite{Wiesendanger}.
Surely, the fabrication of classical logic devices based on quantum dots
arrays will require more advanced technology. However it would be prudent to
begin thinking now about the grounding in theory.

Let us discuss more closely the spin gates proposed by Bandyopadhyay
{\it et al.} \cite{Bandyo} and later investigated theoretically by Molotkov
and Nazin \cite{Molotkov} and by Krasheninnikov and Openov \cite{Openov}.
Such gates are assumed to consist of a number of tunnel-coupled quantum dots
fabricated on a solid surface. The quantum dots play a role of potential
wells for electrons. Each dot is supposed to have a single size-quantized
energy level and there is, on the average, one electron per dot (the total
number of electrons in the array of quantum dots can be controlled by
adjusting the substrate voltage). In a spin gate, the stored information is
determined by the spin configuration of the {\it ground state} of the gate,
i.e., by ground-state averages $\langle\hat{S}_{zi}\rangle$ of spin
projection operators $\hat{S}_{zi}$ at some dots ($i$ is the number of a dot
in the array).

Each spin gate has input and output dots. The input dots serve for writing
the information to the gate through appropriate orientation of their electron
spins (e.g., by the action of the local magnetic field generated
by a magnetic STM tip). The external influence changes the Hamiltonian of the
gate. The new ground state (and its spin configuration) is different from the
initial one since there are electron interactions within (Coulomb repulsion)
and between (tunneling and Coulomb repulsion) dots in the gate. The values
of $\langle\hat{S}_{zi}\rangle$ at the output dots represent the result of
calculations. The number of input and output dots in the gate depends on a
particular logical function realized by the gate. For example, the AND gate
has two inputs and one output, {\it etc}.

The "ground state computing" is the key idea for implementation of classical
Boolean logic on a nanometer scale (we note that this idea was used not only
for spin gates, but also for other kinds of quantum-dots-based gates, see,
e.g., \cite{Lent,Nomoto}). However, the concept of ground state computing has
some essential drawbacks. The first one stems from the fact that the logical
variables 1 and 0 are associated not with certain {\it pure} quantum states,
but with some quantum-mechanical ground-state {\it averages}. For a
particular case of spin gates, the logical 1 (0) at the dot $i$ corresponds
to a positive (negative) value of $\langle\hat{S}_{zi}\rangle$ or, more
precisely, to $\langle\hat{S}_{zi}\rangle \ge S_t/2$
($\langle\hat{S}_{zi}\rangle \le -S_t/2$), where $S_t$ is the "threshold
value", $0<S_t\leq 1$ \cite{Molotkov,Openov}. We note, however, that in a
complex quantum-mechanical system, the electron spin at any quantum dot is
never directed strictly "up" or "down", so the value of $S_t$ cannot be taken
equal to unity. Theoretical analysis of several logical spin gates (NOT, AND,
OR, NAND, NOR, XOR, NXOR, and half-adder) within the framework of spin-1/2
Heisenberg model has shown that for most of those gates the truth tables
(which establish a one-to-one correspondence between the values of input and
output bits) can be realized at sufficiently low values of $S_t$ only
($S_t=0.1$, or even $S_t=0.05$ for half-adder) \cite{Molotkov}. But the
value of $S_t$ should not be too low. Indeed, a low value of $S_t$ does not
imply that {\it average} projection of electron spin
$\langle\hat{S}_{zi}\rangle$ at the dot $i$ has low absolute value (so that
it is hardly seen by a reading device). In fact, since an electron spin is a
quantum-mechanical entity, we can detect only its "up" or "down" directions
at any measurement. A low absolute value of $\langle\hat{S}_{zi}\rangle$ just
means that a probability $p_{\uparrow}$ to find an electron in the state with
spin "up" doesn't differ much from a probability $p_{\downarrow}$ to find an
electron in the state with spin "down". But if so, we will not be able to
draw a definite conclusion about the {\it sign} of
$\langle\hat{S}_{zi}\rangle$ upon one particular measurement. In other words,
there will be a high {\it error probability} $P_{err}$, i.e., by a
definition, a high probability to measure $S_{zi}=-1/2$ while
$\langle\hat{S}_{zi}\rangle > 0$, or to measure $S_{zi}=1/2$ while
$\langle\hat{S}_{zi}\rangle < 0$ ($P_{err}\approx 0.5$ if $S_t<<1$). This is
detrimental for the computational efficiency.

For some simplest spin gates the issue of high error probability, related
to low values of $|\langle\hat{S}_{zi}\rangle|$, can be resolved by applying
a suitably directed uniform external magnetic field to the whole gate
\cite{Openov}. In particular, a fundamental possibility to reach
ground-state averages $|\langle\hat{S}_{zi}\rangle| > 0.475$ at input and
output dots of NOT-AND and NOT-OR three-dot spin gates has been demonstrated
theoretically in \cite{Openov}. Such a large value of
$|\langle\hat{S}_{zi}\rangle|$ corresponds to $P_{err} < 0.05$. However,
apart from the problem of low $|\langle\hat{S}_{zi}\rangle|$, the concept of
ground state computing has one more serious limitation.

The point is that under the influence of external source on input dots of a
spin gate, the relaxation of electronic subsystem to a new ground state can
take place only through energy dissipation. The rate at which dissipation
processes occur determines the speed of operation. However, whereas in
two-dimensional nanostructures (quantum wells) the prevailing relaxation
process is the emission of a longitudinal optical phonon with high relaxation
rates of the order of $10^{12}$ s$^{-1}$, this process is forbidden in
quantum dots because of the discrete nature of energy spectrum. Nomoto
{\it et al.} \cite{Nomoto} have shown that in semiconductor-based quantum
dots, the emission of longitudinal acoustic phonon is the main contribution
to the energy dissipation process. The relaxation rate is $(10^6\div 10^9)$
s$^{-1}$, depending on geometrical parameters of an array of quantum dots
\cite{Nomoto}. The operation speed less than $10^9$ s$^{-1}$ is obviously
too low for quantum-dots-based spin gates to be seriously considered as logic
devices of new generation. Though some other scattering mechanisms (such as
interface-phonon scattering) may also contribute to the energy dissipation in
arrays of quantum dots \cite{Nomoto}, it is unlikely that they will cause
the relaxation rate to increase significantly. Hence, it is reasonable to
search for some other principles of spin gates operation, beyond the concept
of ground state computing.

Recently Bandyopadhyay and Roychowdhury \cite{Bandyo2} have studied
theoretically a possibility of {\it non-dissipative} operation of the
simplest two-dot spin gate NOT (inverter). They showed that under the
influence of local external magnetic field, the logical function can be
realized at the stage of coherent temporal evolution of electronic subsystem
in time less than $10^{-11}$ s, i.e., well before the relaxation to a new
ground state takes place at the sacrifice of inelastic interactions.
Moreover, it was shown \cite{Bandyo2} that there exists an optimal input
signal energy to achieve a {\it complete switching} of the inverter, i.e.,
to realize spin configurations with saturated spin projections
($\langle \hat{S}_{zi}\rangle$ = 1/2 or -1/2) at input and output dots. The
results obtained in \cite{Bandyo2} are very encouraging. However, the authors
of Ref.\cite{Bandyo2} used the spin-1/2 antiferromagnetic Heisenberg model to
describe the correlated electrons in quantum dots constituting the inverter.
As is known \cite{Izyumov}, the Heisenberg model is just a limiting case of
the half-filled Hubbard model at $U>>V$, where $U$ is the intra-dot Coulomb
repulsion energy and $V$ is the inter-dot tunneling energy. Since it is not
clear {\it a priori} to what extent the Heisenberg model can describe the
actual experimental situation, it is instructive to make use of more
realistic Hubbard model, without any constraints on the ratio $U/V$.

In this paper we study the non-dissipative dynamics of the inverter within
the framework of the Hubbard model (preliminary results have been presented
in \cite{Openov2}). We explore a broad range of $U$ and $V$ values which
physically correspond to various size of quantum dots and to various distance
between them (the Heisenberg model as a limiting case at $U>>V$ is also
considered since it allows for a simple analytical solution). We show that
complete switching of the inverter can be realized at unrealistic values of
$U/V=0$ and $U/V=\infty$ only. Nevertheless, it turns out that at {\it any}
value of $U/V$ one can reach averages $|\langle\hat{S}_{zi}\rangle| > 0.45$
at input and output dots in a very short operation time of the order of
$10^{-11}$ s. Such large values of $|\langle\hat{S}_{zi}\rangle|$ correspond
to the error probability $P_{err}<0.1$ which is sufficiently low for
implementation of logic function.

The paper is organized as follows. The Hubbard and Heisenberg models for the
two-dot spin gate NOT (inverter) are described in Sec.2. The ground state
computing is briefly outlined in Sec.3 for completeness sake. The temporal
evolution of electronic subsystem, once the local external magnetic field is
applied to the input dot, is studied in Sec.4, with special emphasis on
time-dependent quantum-mechanical spin averages at input and output dots. We
also analyze the dynamics of the inverter upon removing the input signal. The
results obtained are discussed in Sec.5 from the viewpoint of prospects for
non-dissipative computing in spin gates based on quantum dots. In Sec.6,
concluding remarks are given.

\vskip 6mm
\centerline{\bf 2. Theoretical models}

\vskip 2mm

The inverter consists of two closely spaced quantum dots ($A$ and $B$)
occupied by two electrons \cite{Bandyo,Molotkov,Bandyo2}. One of two dots
(say, the dot $A$) serves for writing the input signal to the gate by the
action of the local external magnetic field $H_A$. The second dot ($B$) is
the output.

\vskip 4mm

\centerline{\bf A. Hubbard model}

\vskip 2mm

Under the assumption that each dot has only one size-quantized energy level,
electron-dot interactions can be parametrized by one-electron Hartree-Fock
on-site energies $\epsilon_A$, $\epsilon_B$, and the energy $V$ of electron
tunneling between the dots. Electron-electron interactions are specified by
the energies $U_{AA}$, $U_{BB}$ and $U_{AB}$ of Coulomb repulsion between two
electrons residing at the same dot or at different dots. We ignore the
exchange Coulomb interaction between the dots \cite{Molotkov} since, first,
its characteristic energy is generally much smaller than $U_{ij}$ ($i,j=A,B$)
\cite{Hubbard} and, second, the antiferromagnetic exchange naturally arises
in the limiting case $V<<U_{ij}$ which corresponds to the reduction of the
Hubbard model to the Heisenberg one \cite{Izyumov}, see below. The presence
of the (parallel to $z$-axis) local external magnetic field $H_A$ at the dot
$A$ leads to the change in the total energy by $\mu H_A$, where $\mu$ is
$z$-component of the magnetic moment at the dot $A$. In the site
representation, the resulting Hubbard-like Hamiltonian has the form
\begin{eqnarray}
\hat{H}&=&\epsilon_A\sum_{\sigma}\hat{n}_{A\sigma}+
\epsilon_B\sum_{\sigma}\hat{n}_{B\sigma}-
V\sum_{\sigma}(\hat{a}^{+}_{A\sigma}\hat{a}^{}_{B\sigma}+
\hat{a}^{+}_{B\sigma}\hat{a}^{}_{A\sigma})+ \nonumber \\
&&U_{AA}\hat{n}_{A\uparrow}\hat{n}_{A\downarrow}+
U_{BB}\hat{n}_{B\uparrow}\hat{n}_{B\downarrow}+
U_{AB}\sum_{\sigma,\sigma^{\prime}}%
\hat{n}_{A\sigma}\hat{n}_{B\sigma^{\prime}}-g\mu_B H_A \hat{S}_{zA},
\label{Hamiltonian1}
\end{eqnarray}
where $\hat{a}^{}_{i\sigma}$ $(\hat{a}^{+}_{i\sigma})$ are the operators for
annihilation (creation) of electron with the spin projection
$\sigma=~\uparrow$ or $\downarrow$ at the dot $i=A$ or $B$,
$\hat{n}_{i\sigma}=\hat{a}^{+}_{i\sigma}\hat{a}^{}_{i\sigma}$
are the electron number operators,
$\hat{S}_{zi}=(\hat{n}_{i\uparrow}-\hat{n}_{i\downarrow})/2$ is the operator
of spin projection at the dot $i$, $g$ is the Lande factor, $\mu_B$ is the
Bohr magneton. The external magnetic field is directed along the magnetic
moment of electron having $\sigma=~\uparrow$.

Now let us slightly simplify the Hamiltonian (\ref{Hamiltonian1}), keeping
all the key terms in it. First, we assume that the dots $A$ and $B$ are
identical, so that $\epsilon_A=\epsilon_B=\epsilon_0$ and $U_{AA}=U_{BB}=U_0$.
Second, we suppose that the total number of electrons in the inverter
$N=\hat{n}_{A\uparrow}+\hat{n}_{A\downarrow}+\hat{n}_{B\uparrow}+
\hat{n}_{B\downarrow}$ equals to 2 (one electron per dot, i.e., the
half-filled band) and remains unchanged during the operation
\cite{Molotkov,Bandyo2} at least till the reading of the result from the
output dot. Then, by making use of the identity (see, e.g., \cite{Nomoto})
$$2\sum_{\sigma,\sigma^{\prime}}\hat{n}_{A\sigma}\hat{n}_{B\sigma^{\prime}}+
2\hat{n}_{A\uparrow}\hat{n}_{A\downarrow}+
2\hat{n}_{B\uparrow}\hat{n}_{B\downarrow}=N(N-1),$$
we obtain
\begin{eqnarray}
\hat{H}&=&-V\sum_{\sigma}(\hat{a}^{+}_{A\sigma}\hat{a}^{}_{B\sigma}+
\hat{a}^{+}_{B\sigma}\hat{a}^{}_{A\sigma})+
(U_0-U_{AB})(\hat{n}_{A\uparrow}\hat{n}_{A\downarrow}+
\hat{n}_{B\uparrow}\hat{n}_{B\downarrow})- \nonumber \\
&&\frac{g\mu_B H_A}{2}\sum_{\sigma}\hat{n}_{A\sigma}sign(\sigma)+
\epsilon_0 N + U_{AB}\frac{N(N-1)}{2},
\label{Hamiltonian2}
\end{eqnarray}
where $sign(\sigma)$ = +1 and -1 for $\sigma=~\uparrow$ and $\downarrow$
respectively. Finally, omitting two last non-operator terms from the
Hamiltonian (\ref{Hamiltonian2}), we arrive at the expression
\begin{equation}
\hat{H}=-V\sum_{\sigma}(\hat{a}^{+}_{A\sigma}\hat{a}^{}_{B\sigma}+
\hat{a}^{+}_{B\sigma}\hat{a}^{}_{A\sigma})+
U(\hat{n}_{A\uparrow}\hat{n}_{A\downarrow}+
\hat{n}_{B\uparrow}\hat{n}_{B\downarrow})-
H\sum_{\sigma}\hat{n}_{A\sigma}sign(\sigma),
\label{Hamiltonian3}
\end{equation}
where we have introduced the renormalized energy of intra-dot Coulomb
repulsion, $U=U_0-U_{AB}$, and designated the input signal energy
$H=g\mu_BH_A/2$. We note that generally $U>0$ since $U_0>U_{AB}$
\cite{Nomoto}.

The complete orthonormal set of inverter eigenstates is formed by the
two-electron basis states
\begin{equation}
|1\rangle=|\uparrow,\downarrow\rangle,~
|2\rangle=|\downarrow,\uparrow\rangle,~
|3\rangle=|\uparrow\downarrow,0\rangle,~
|4\rangle=|0,\uparrow\downarrow\rangle,~
|5\rangle=|\uparrow,\uparrow\rangle,~
|6\rangle=|\downarrow,\downarrow\rangle,
\label{basis}
\end{equation}
where, e.g., the notation $|\uparrow,\downarrow\rangle$ denotes the state
with up-spin electron at the dot $A$ and down-spin electron at the dot $B$,
the notation $|\uparrow\downarrow,0\rangle$ denotes the state with two
(up-spin and down-spin) electrons at the dot $A$ and no electrons at the dot
$B$, {\it etc}. We stress that the spin projection on either of the two
quantum dots is zero in the basis states $|3\rangle$ and $|4\rangle$. In this
respect the NOT gate based on two {\it quantum dots} differs in general from
the "spatially extended NOT gate" based on two {\it spins} \cite{Mozyrsky}.
In fact, the latter corresponds to the limiting case $U>>V$ of the
Hamiltonian (\ref{Hamiltonian3}) and may be described within the framework of
the Heisenberg model, see below.

In the virgin state (i.e., at $H=0$), the time-dependent electron wave
function $\Psi_0(t)$ is the solution of the Schr\"odinger equation
\begin{equation}
i\hbar\partial\Psi_0(t)/\partial t=\hat{H}\Psi_0(t)
\label{Schrodinger_t}
\end{equation}
with the Hamiltonian (\ref{Hamiltonian3}) taken at $H=0$ and can be
represented as
\begin{equation}
\Psi_0(t)=\sum_{k=1}^{6}A_k^{(0)}\Psi_k^{(0)}\exp(-iE_k^{(0)}t/\hbar),
\label{Psi0(t)}
\end{equation}
where the coefficients $A_k^{(0)}$ do not depend on time, and $\Psi_k^{(0)}$
and $E_k^{(0)}$ $(k=1\div 6)$ are eigenvectors and eigenenergies of the
stationary Schr\"odinger equation
\begin{equation}
\hat{H}\Psi_k^{(0)}=E_k^{(0)}\Psi_k^{(0)}
\label{Schrodinger0}
\end{equation}
They are as follows:
\begin{eqnarray}
&&\Psi_1^{(0)}=\frac{1}{\sqrt{2}}(|1\rangle-|2\rangle),~
\Psi_2^{(0)}=\frac{1}{\sqrt{2}}(|3\rangle-|4\rangle), \nonumber \\
&&\Psi_3^{(0)}=\frac{1}{2}\sqrt{1+\frac{U}{\sqrt{U^2+16V^2}}}\left(|1\rangle+
|2\rangle+\frac{\sqrt{U^2+16V^2}-U}{4V}|3\rangle+
\frac{\sqrt{U^2+16V^2}-U}{4V}|4\rangle\right), \nonumber \\
&&\Psi_4^{(0)}=\frac{1}{2}\sqrt{1-\frac{U}{\sqrt{U^2+16V^2}}}\left(|1\rangle+
|2\rangle-\frac{\sqrt{U^2+16V^2}+U}{4V}|3\rangle-
\frac{\sqrt{U^2+16V^2}+U}{4V}|4\rangle\right), \nonumber \\
&&\Psi_5^{(0)}=|5\rangle,~\Psi_6^{(0)}=|6\rangle,
\label{Psi0_k}
\end{eqnarray}
and
\begin{equation}
E_1^{(0)}=0,~E_2^{(0)}=U,~E_3^{(0)}=-\frac{\sqrt{U^2+16V^2}-U}{2},~
E_4^{(0)}=\frac{\sqrt{U^2+16V^2}+U}{2},~E_5^{(0)}=0,~E_6^{(0)}=0.
\label{E0_k}
\end{equation}

Generally, the coefficients $A_k^{(0)}$ in (\ref{Psi0(t)}) can take any
values which satisfy the normalization condition
\begin{equation}
\sum_{k=1}^{6}|A_k^{(0)}|^2=1.
\label{norma0}
\end{equation}
In other words, the electronic subsystem can exist in a rather complex
superposition of quantum eigenstates $\Psi_k^{(0)}$.

In the presence of the input signal at the dot $A$, the time-dependent wave
function $\Psi(t)$ satisfies the Schr\"odinger equation (\ref{Schrodinger_t})
with the Hamiltonian (\ref{Hamiltonian3}), and can be expanded as
\begin{equation}
\Psi(t)=\sum_{k=1}^{6}A_k\Psi_k\exp(-iE_kt/\hbar),
\label{Psi(t)}
\end{equation}
where $\Psi_k$ and $E_k$ $(k=1-6)$ are eigenvectors and eigenenergies of
the stationary Schr\"odinger equation
\begin{equation}
\hat{H}\Psi_k=E_k\Psi_k,
\label{Schrodinger}
\end{equation}
and the time-independent coefficients $A_k$ satisfy the normalization
condition
\begin{equation}
\sum_{k=1}^{6}|A_k|^2=1.
\label{norma}
\end{equation}

At arbitrary values of $U$, $V$, and $H$ the eigenvalue equation
(\ref{Schrodinger}) reduces to the algebraic equation of the third power in
$E_k$. The resulting expressions are too cumbersome for analysis, so it is
more convenient to solve the equation (\ref{Schrodinger}) numerically.
However, it is useful for the following discussion to consider a special case
$U=0$ which physically corresponds to $U<<V$ (closely-spaced large-sized dots
\cite{Nomoto}). At $U=0$ and at arbitrary values of $V$ and $H$ we have for
eigenvectors and eigenenergies of the equation (\ref{Schrodinger}) rather
simple analytical expressions:
\begin{eqnarray}
&&\Psi_1=\sqrt{\frac{2V^2}{H^2+4V^2}}\left(|1\rangle-|2\rangle-
\frac{H}{2V}|3\rangle-\frac{H}{2V}|4\rangle\right),~
\Psi_2=\frac{1}{\sqrt{2}}(|3\rangle-|4\rangle), \nonumber \\
&&\Psi_3=\frac{\sqrt{H^2+4V^2}+H}{2\sqrt{H^2+4V^2}}\left(|1\rangle+
\frac{\sqrt{H^2+4V^2}-H}{\sqrt{H^2+4V^2}+H}|2\rangle+
\frac{\sqrt{H^2+4V^2}-H}{2V}|3\rangle+
\frac{\sqrt{H^2+4V^2}-H}{2V}|4\rangle\right), \nonumber \\
&&\Psi_4=\frac{\sqrt{H^2+4V^2}-H}{2\sqrt{H^2+4V^2}}\left(|1\rangle+
\frac{\sqrt{H^2+4V^2}+H}{\sqrt{H^2+4V^2}-H}|2\rangle-
\frac{\sqrt{H^2+4V^2}+H}{2V}|3\rangle-
\frac{\sqrt{H^2+4V^2}+H}{2V}|4\rangle\right), \nonumber \\
&&\Psi_5=|5\rangle,~\Psi_6=|6\rangle,
\label{Psi_k}
\end{eqnarray}
and
\begin{equation}
E_1=0,~E_2=0,~E_3=-\sqrt{H^2+4V^2},~E_4=\sqrt{H^2+4V^2},~E_5=-H,~E_6=H.
\label{E_k}
\end{equation}
One can see that expressions (\ref{Psi_k}) and (\ref{E_k}) taken at $H=0$
coincide with expressions (\ref{Psi0_k}) and (\ref{E0_k}) taken at $U=0$
respectively .

\vskip 4mm

\centerline{\bf B. Heisenberg model}

\vskip 2mm

In the limiting case $U>>V$ (widely-spaced small-sized dots \cite{Nomoto}),
the half-filled Hubbard model reduces to the spin-1/2 antiferromagnetic
Heisenberg model \cite{Izyumov}. The corresponding Hamiltonian for the
inverter has the form \cite{Molotkov,Bandyo2}:
\begin{equation}
\hat{H}=J\sum_{\alpha}\hat{\sigma}_{\alpha A}\hat{\sigma}_{\alpha B}-
H\sum_{\sigma}\hat{n}_{A\sigma}sign(\sigma),
\label{Heisenberg}
\end{equation}
where $\hat{\sigma}_{\alpha i}$ ($\alpha=x,y,z$) are the Pauli matrices
describing the electron at the dot $i=$ $A$ or $B$ and $J=V^2/U$ is the
energy of exchange interaction between two electrons residing at dots $A$ and
$B$. We have omitted the term $-J(\hat{n}_{A\uparrow}+\hat{n}_{A\downarrow})%
(\hat{n}_{B\uparrow}+\hat{n}_{B\downarrow})$ \cite{Izyumov} from
(\ref{Heisenberg}) since in the Heisenberg model the complete orthonormal set
of inverter eigenstates doesn't include the two-electron states
$|3\rangle=|\uparrow\downarrow,0\rangle$ and
$|4\rangle=|0,\uparrow\downarrow\rangle$, see (\ref{basis}), so that each dot
is occupied by one electron, and hence
$(\hat{n}_{A\uparrow}+\hat{n}_{A\downarrow})%
(\hat{n}_{B\uparrow}+\hat{n}_{B\downarrow})=1$ is just a $c$-number.

The Heisenberg model for the inverter allows for analytical solution at any
values of $J$ and $H$ \cite{Molotkov,Bandyo2} (in what follows we shall make
use of this fact to analize the case $U>>V$). The eigenvectors and
eigenenergies of the stationary Schr\"odinger equation with the Hamiltonian
(\ref{Heisenberg}) are as follows:
\begin{eqnarray}
&&\Psi_1=\sqrt{\frac{1}{2}\left(1+\frac{H}{\sqrt{H^2+4J^2}}\right)}|1\rangle-
\sqrt{\frac{1}{2}\left(1-\frac{H}{\sqrt{H^2+4J^2}}\right)}|2\rangle,
\nonumber \\
&&\Psi_2=\sqrt{\frac{1}{2}\left(1-\frac{H}{\sqrt{H^2+4J^2}}\right)}|1\rangle+
\sqrt{\frac{1}{2}\left(1+\frac{H}{\sqrt{H^2+4J^2}}\right)}|2\rangle,
\nonumber \\
&&\Psi_3=|5\rangle,~\Psi_4=|6\rangle,
\label{Psi_k_Heisenberg}
\end{eqnarray}
and
\begin{equation}
E_1=-J-\sqrt{H^2+4J^2},~E_2=-J+\sqrt{H^2+4J^2},~E_3=J-H,~E_4=J+H.
\label{E_k_Heisenberg}
\end{equation}

The time-dependent wave function $\Psi(t)$ has the form (\ref{Psi(t)}) with
$\Psi_k$ and $E_k$ given by (\ref{Psi_k_Heisenberg}) and
(\ref{E_k_Heisenberg}) respectively except that the number of inverter
eigenstates in the Heisenberg model is less by two than in the Hubbard one,
so that the summation over $k$ in (\ref{Psi(t)}) is from $k=1$ up to $k=4$.

\vskip 6mm

\centerline{\bf 3. Ground state computing}

\vskip 2mm

The truth table of the logical gate NOT is very simple, the output bit $B$
being just the inverse of the input bit $A$, i.e., $B=\overline{{A}}$
($B=1$ if $A=0$ and $B=0$ if $A=1$). Let us see how this truth table is
realized if the input and output bits are encoded in {\it ground state}
averages $\langle\hat{S}_{zA}\rangle$ and $\langle\hat{S}_{zB}\rangle$
respectively. We shall ascribe the bit 1 (0) at the dot $i$ to the positive
(negative) value of $\langle\hat{S}_{zi}\rangle$, with the constraint
$|\langle\hat{S}_{zi}\rangle|\ge S_t/2$, where $0<S_t\le 1$ \cite{Molotkov,%
Openov}.

As discussed in the Introduction, one wishes to have $S_t$ as large as
possible since the greater is $S_t$, the lower is the error probability
$P_{err}$, i.e., the probability to read the "wrong" signal $S_{zi}$ at the
dot $i$ (e.g., to measure $S_{zi}=-1/2$ while
$\langle\hat{S}_{zi}\rangle>0$). The physical reason for a possibility of
such an error in the device operation is the entangled structure of the
ground state wave function (the latter cannot be factored as
$|\uparrow\rangle _A|\downarrow\rangle _B$ or
$|\downarrow\rangle _A|\uparrow\rangle_B$). It is straightforward to find the
upper limit of $P_{err}$ as a function of $S_t$. For the sake of
definiteness, let us consider the case $\langle \hat{S}_{zA}\rangle\ge S_t/2$,
$\langle \hat{S}_{zB}\rangle\le -S_t/2$ which corresponds to $H>0$. Then
$P_{err}=1-p_{\uparrow\downarrow}$ where $p_{\uparrow\downarrow}$ is the
quantum-mechanical probability to find up-spin electron at the dot $A$ and
down-spin electron at the dot $B$. Since, by a definition,
$\langle \hat{S}_{zA}\rangle=(p_{\uparrow\downarrow}-
p_{\downarrow\uparrow})/2$, one has generally
$p_{\uparrow\downarrow}-p_{\downarrow\uparrow}\ge S_t$, and hence
$P_{err}\le 1-S_t$ (in the special case $U>>V$, the ground state eigenvector
is composed of states $|\uparrow,\downarrow\rangle$ and
$|\downarrow,\uparrow\rangle$ only, see (\ref{basis}),
(\ref{Psi_k_Heisenberg}), (\ref{E_k_Heisenberg}), and one has
$p_{\uparrow\downarrow}+p_{\downarrow\uparrow}=1$,
$p_{\uparrow\downarrow}\ge (1+S_t)/2$, and $P_{err} \le (1-S_t)/2$). Note
that $P_{err}=0$ in the ideal case $S_t=1$ only, i.e., in the case of
saturated spin projections $|\langle\hat{S}_{zi}\rangle|=1/2$ at both dots
$i=A$ and $B$.

It follows from (\ref{Psi0_k}) and (\ref{E0_k}) that at zero input signal
$H=0$ the ground state of the inverter at any value of the ratio $U/V$ is the
entangled state $\Psi_3^{(0)}$ with zero magnetic moments at both dots $A$
and $B$ since $\langle\Psi_3^{(0)}|\hat{S}_{zA}|\Psi_3^{(0)}\rangle=
\langle\Psi_3^{(0)}|\hat{S}_{zB}|\Psi_3^{(0)}\rangle=0$. At $H\neq 0$, the
ground state averages of spin projections at the input and output dots become
nonzero. They are equal in magnitude and opposite in sign,
$\langle \hat{S}_{zA}\rangle=-\langle \hat{S}_{zB}\rangle$, just as required
by the truth table of the inverter. Making use of (\ref{Psi_k}) and
(\ref{E_k}), one has
\begin{equation}
\langle \hat{S}_{zA}\rangle=\frac{H}{2\sqrt{H^2+4V^2}},~~U<<V,
\label{SzA-U0}
\end{equation}
while
\begin{equation}
\langle \hat{S}_{zA}\rangle=\frac{H}{2\sqrt{H^2+4J^2}},~~U>>V
\label{SzA-Uinf}
\end{equation}
as follows from (\ref{Psi_k_Heisenberg}) and (\ref{E_k_Heisenberg}). At
arbitrary value of the ratio $U/V$, the dependence of
$\langle \hat{S}_{zA}\rangle$ on $H$ may be approximated as
$\langle \hat{S}_{zA}\rangle=H/2\sqrt{H^2+4W^2}$ where $W=Vf(U/V)$, and the
function $f(x)$ is shown in Fig. 1 ($f(0)=1$; $f(x)=1/x$ at $x>>1$). Thus,
$\langle \hat{S}_{zA}\rangle$ increases monotonically with $H$ from
$\langle \hat{S}_{zA}\rangle=0$ at $H=0$ and asimptotically approaches its
maximum value $\langle \hat{S}_{zA}\rangle=1/2$ at $H>>W$. The error
probability is close to 0.75 (for $U<<V$) or 0.5 (for $U>>V$) at $H<<W$ and
decreases with $H$ as $P_{err}\approx \alpha W^2/H^2$ at $H>>W$
($\alpha$ = 2 and 1 for $U<<V$ and $U>>V$ respectively).

In order to minimize $P_{err}$ one should have the input signal energy as
high as possible. However this cannot be easily realized experimentally since
realistic values of $H_A\leq 1$ T correspond to quite low values of
$H\leq 0.1$ meV, while $H$ should be much greater than the characteristic
energy $W$. On the other hand, the energy $W$ should exceed considerably the
thermal energy $k_B T$ for stable operation \cite{Bandyo2}, so the operating
temperature should be restricted to at least 100 mK.

Aside from the problem of $P_{err}$ reduction, the speed of device operation
is limited by the finite rate of relaxation of electronic subsystem to a new
ground state upon applying the input signal (see the Introduction). The
relaxation rate is determined by dissipation processes, so the dissipation is
a necessary attribute of the ground state computing in the array of quantum
dots. In what follows we study a possibility to perform a logical operation
NOT {\it in the absence of dissipation}, at the stage of non-dissipative
temporal evolution of electronic subsystem.

\vskip 6mm

\centerline{\bf 4. Non-dissipative computing}

\vskip 2mm

Our purpose is to calculate the quantum-mechanical averages
\begin{equation}
S_{zA}(t)=-S_{zB}(t)=\langle \Psi (t)|\hat{S}_{zA}|\Psi (t)\rangle
\label{SzA(t)}
\end{equation}
at $t\geq 0$, i.e., upon applying the input signal to the dot $A$. For
definiteness, we shall consider the case $H_A>0$ (i.e., $H>0$). We shall
also assume that at $t\leq 0$ (in the absence of external signal) the system
is in its ground state, so that the time-dependent wave function at $t\leq 0$
is $\Psi_0(t)=\Psi_3^{(0)}\exp(-iE_3^{(0)}t/\hbar)$, see (\ref{Psi0_k}),
(\ref{E0_k}), and hence the coefficients $A_k^{(0)}$ in the expansion of
$\Psi_0(t)$ (\ref{Psi0(t)}) have the values
$A_1^{(0)}=A_2^{(0)}=A_4^{(0)}=A_5^{(0)}=A_6^{(0)}=0$, $A_3^{(0)}=1$.

The coefficients $A_k$ in the expansion of the time-dependent wave function
$\Psi(t)$ (\ref{Psi(t)}) at $t\geq 0$ should be found from the initial
condition $\Psi(t=0)=\Psi_0(t=0)$. It is convenient to write the eigenvectors
$\Psi_k$ of the stationary Schr\"odinger equation (\ref{Schrodinger}) as
\begin{equation}
\Psi_k=\sum_{n=1}^{6}B_{kn}|n\rangle.
\label{Psi_k_B}
\end{equation}
Then
\begin{equation}
\Psi(t)=\sum_{n=1}^{6}f_n(t)|n\rangle,
\label{Psi(t)_f}
\end{equation}
where
\begin{equation}
f_n(t)=\sum_{k=1}^{6}A_kB_{kn}\exp(-iE_kt/\hbar).
\label{f_n}
\end{equation}
The probability to find the system in the basis state $|n\rangle$ at time $t$
is
\begin{equation}
p_n(t)=|f_n(t)|^2.
\label{p_n}
\end{equation}
Since, as follows from (\ref{basis}), $\langle 1|\hat{S}_{zA}|1\rangle=1/2$,
$\langle 2|\hat{S}_{zA}|2\rangle=-1/2$, $\langle 3|\hat{S}_{zA}|3\rangle=%
\langle 4|\hat{S}_{zA}|4\rangle=0$, $\langle 5|\hat{S}_{zA}|5\rangle=1/2$,
$\langle 6|\hat{S}_{zA}|6\rangle=-1/2$, one has from (\ref{SzA(t)}),
(\ref{Psi(t)_f}), (\ref{p_n}):
\begin{equation}
S_{zA}(t)=-S_{zB}(t)=\frac{p_1(t)-p_2(t)+p_5(t)-p_6(t)}{2}.
\label{SzA(t)_2}
\end{equation}

First we consider the limiting case $U<<V$ \cite{Openov2} (the weak coupling
limit of the Hubbard model). Setting $U=0$, we have rather simple equations
for the probabilities $p_n(t)$:
\begin{eqnarray}
&&p_1(t)=\frac{1}{4}\left(1+\frac{4HV}%
{H^2+4V^2}\sin^2(\omega t/2)\right)^2 ,\nonumber \\
&&p_2(t)=\frac{1}{4}\left(1-\frac{4HV}%
{H^2+4V^2}\sin^2(\omega t/2)\right)^2 ,\nonumber \\
&&p_3(t)=p_4(t)=\frac{1}{4}\left(1-\frac{16H^2V^2}%
{(H^2+4V^2)^2}\sin^4(\omega t/2)\right) ,\nonumber \\
&&p_5(t)=p_6(t)=0 ,
\label{p_n_U0}
\end{eqnarray}
where
\begin{equation}
\omega=\sqrt{H^2+4V^2}/\hbar .
\label{omega_1}
\end{equation}
From (\ref{SzA(t)_2}) and
(\ref{p_n_U0}) it is straightforward to find that
\begin{equation}
S_{zA}(t)=-S_{zB}(t)=\frac{2HV}{H^2+4V^2}\sin^2(\omega t/2).
\label{SzA(t)_U0}
\end{equation}

It follows from (\ref{SzA(t)_U0}) that at any positive values of $H$ and $V$
the function $S_{zA}(t)$ is non-negative and oscillates in time with the
frequency $\omega$ (given by (\ref{omega_1})) and the amplitude
\begin{equation}
S_{zA}(t_0)=\frac{2HV}{H^2+4V^2},
\label{SzA(t_0)_U0}
\end{equation}
where $t_0$ stands for the time of the first peak of $S_{zA}(t)$,
\begin{equation}
t_0=\frac{\pi}{\omega}=\frac{\pi\hbar}{\sqrt{H^2+4V^2}}.
\label{t_0_U0}
\end{equation}
The curve $S_{zA}(t)$ is shown in Fig. 2 for several different values of
$H/V$. From Fig. 2 and (\ref{omega_1}), (\ref{t_0_U0}) one can see that the
oscillation frequency $\omega$ monotonically increases (i.e., the time $t_0$
decreases) with $H$, while the amplitude $S_{zA}(t_0)$ first increases with
$H$, reaches the maximum value 1/2 at $H=2V$ and then decreases, being in
inverse proportion to $H$ at $H>>V$. Thus, a {\it complete} switching,
$S_{zA}(t_0)=1/2$ and $S_{zB}(t_0)=-1/2$, is achieved at $H/V=2$ and
$t_0=\pi\hbar/2\sqrt{2}V$.

It should be stressed that though the function $S_{zA}(t)$, being periodic in
time, has its peak value at $t_k=\pi/\omega+2\pi k/\omega$ ($k$ is an
integer), here and below we are interested in the {\it lowest} possible value
of $t_k$ (i.e., $t_0$) since we want not only to reach the maximum
permissible value of $S_{zA}$, but to do it in as short as possible switching
time.

Now let us turn to the opposite limiting case, $U>>V$ (the strong coupling
limit of the Hubbard model). This case may be described within the framework
of the Heisenberg model (\ref{Heisenberg}). It has been considered by
Bandyopadhyay and Roychowdhury in \cite{Bandyo2}. For the sake of
completeness, we briefly sketch the results obtained in \cite{Bandyo2}.

As we have noted above, in the Heisenberg model, the complete set of inverter
eigenstates doesn't include the states
$|3\rangle=|\uparrow\downarrow,0\rangle$ and
$|4\rangle=|0,\uparrow\downarrow\rangle$, see (\ref{basis}),
(\ref{Psi_k_Heisenberg}). For the probabilities $p_n(t)$ to find the system
in the basis states $|n\rangle$ = $|1\rangle$, $|2\rangle$, $|5\rangle$ and
$|6\rangle$ one has
\begin{eqnarray}
&&p_1(t)=\frac{1}{2}\left(1+\frac{4HJ}%
{H^2+4J^2}\sin^2(\omega t/2)\right) ,\nonumber \\
&&p_2(t)=\frac{1}{2}\left(1-\frac{4HJ}%
{H^2+4J^2}\sin^2(\omega t/2)\right) ,\nonumber \\
&&p_5(t)=p_6(t)=0~,
\label{p_n_t-J}
\end{eqnarray}
where
\begin{equation}
\omega=2\sqrt{H^2+4J^2}/\hbar
\label{omega_2}
\end{equation}
(we recall that $J=V^2/U<<V$ is the energy of antiferromagnetic exchange).
From (\ref{SzA(t)_2}) and (\ref{p_n_t-J}) one has
\begin{equation}
S_{zA}(t)=-S_{zB}(t)=\frac{2HJ}{H^2+4J^2}\sin^2(\omega t/2).
\label{SzA(t)_t-J}
\end{equation}
One can see from (\ref{SzA(t)_t-J}) that the function $S_{zA}(t)$ at $U>>V$
has the same form as in the case $U<<V$ (\ref{SzA(t)_U0}). The difference
between the two cases is that $V$ in (\ref{omega_1}) and (\ref{SzA(t)_U0}) is
replaced by $J$ in (\ref{omega_2}) and (\ref{SzA(t)_t-J}) respectively, and
that there is an extra factor of 2 in the oscillation frequency in
(\ref{omega_2}) as compared with (\ref{omega_1}). The amplitude of the
function $S_{zA}(t)$ is
\begin{equation}
S_{zA}(t_0)=\frac{2HJ}{H^2+4J^2},
\label{SzA(t_0)_t-J}
\end{equation}
where $t_0$ is the time of the first peak of $S_{zA}(t)$,
\begin{equation}
t_0=\frac{\pi}{\omega}=\frac{\pi\hbar}{2\sqrt{H^2+4J^2}}.
\label{t_0_t-J}
\end{equation}
It follows from (\ref{SzA(t_0)_t-J}) and (\ref{t_0_t-J}) that a
{\it complete} switching, $S_{zA}(t_0)=1/2$ and $S_{zB}(t_0)=-1/2$, is
achieved at $H/J=2$ and $t_0=\pi\hbar/4\sqrt{2}J$.

Thus, having considered two limiting cases of weak ($U<<V$) and strong
($U>>V$) electron coupling, we see that the complete switching of the
inverter can take place in either limit of the Hubbard model providing a
proper choice of the values of the input signal energy $H$ and the switching
time $t_0$. Hence, one may expect that the complete switching can occur also
in the case of intermediate coupling, at an arbitrary ratio of $U$ to $V$
(i.e., at arbitrary values of the size of quantum dots and the distance
between them).

To check this hypothesis, we have calculated numerically the dependencies of
$S_{zA}$ on $t$ at different values of $U/V$ and $H/V$. Several typical
examples of $S_{zA}$ versus $t$ curves are shown in Fig. 3. The switching
time $t_0$ was generally defined as a time of the first peak on the curve
$S_{zA}(t)$, $t_0$ being a function of $U/V$ and $H/V$. As opposed to the
cases of weak and strong coupling considered above, at arbitrary values of
$U/V$ and $H/V$ the function $S_{zA}(t)$ is not periodic in time since it
includes several harmonics with different frequencies and amplitudes. Hence,
in principle, the value of $S_{zA}(t)$ greater than $S_{zA}(t_0)$ can be
achieved at some longer time (see, e.g., Fig. 3b). But this case is of no
interest for us since our purpose is not only to maximize $S_{zA}$, but also
to minimize the switching time.

The curves of $S_{zA}(t_0)$ versus $H/V$ are shown in Fig. 4 for several
values of $U/V$. One can see that an increase in $U/V$ first results in a
decreased height of the maximum $S_{zA}^{max}$ on the $S_{zA}(t_0)$ versus
$H/V$ curve. For $U/V>2$ the value of $S_{zA}^{max}$ increases once more, but
doesn't reach the saturation value 1/2 at finite $U/V$, though
$S_{zA}^{max}\rightarrow 1/2$ if $U/V\rightarrow\infty$ (this corresponds to
the Heisenberg model and agrees with the results obtained in
\cite {Bandyo2}). Fig. 5 shows the dependence of $S_{zA}^{max}$ on $U/V$.
Note that $S_{zA}^{max}\ge 0.45$ at any $U/V$.

We define $H_{opt}$ as the "optimal" value of input signal energy $H$, i.e.,
as the value of $H$ at which the peak value $S_{zA}(t_0)$ reaches its maximum
$S_{zA}^{max}$ at a given $U/V$ (e.g., $H_{opt}=2V$ at $U/V=0$, see
(\ref{SzA(t_0)_U0}); $H_{opt}=2J=2V^2/U$ at $U>>V$, see (\ref{SzA(t_0)_t-J});
$H_{opt}=0.83V$ at $U/V=2$, see Fig. 4). The dependence of $H_{opt}/V$ on
$U/V$ is plotted in Fig. 6. One can see from this figure that $H_{opt}/V$
decreases monotonically with increasing $U/V$. It is remarkable that the
equation $H_{opt}/V=2V/U$ derived analytically in the Heisenberg model,
i.e., for $U>>V$ (\ref{SzA(t_0)_t-J}), is a good approximation of the
numerically calculated curve even at $U\approx V$. It follows from Fig. 6
that the curves $S_{zA}(t)$ shown in Figs. 3a and 3b are for the case
$H>H_{opt}$ while the curve $S_{zA}(t)$ shown in Fig. 3c is for the case
$H=H_{opt}$ (hence, the peak value of $S_{zA}(t_0)$ in Fig. 3c is the maximum
permissible value of $S_{zA}(t)$ at $U/V=1.5$).

An important characteristic of the logical gate is its switching time $t_0$
which depends on $U/V$ and $H/V$ (see above). At a given $U/V$, we define the
"optimal" switching time $t_{opt}$ as the time of the first peak on the curve
$S_{zA}(t)$ at $H=H_{opt}$. Hence, $S_{zA}(t_{opt})$ equals to
$S_{zA}^{max}$, the maximum permissible value of $S_{zA}(t_0)$ at fixed
$U/V$ (see Fig. 4). The dependence of $t_{opt}$ on $U/V$ is shown in Fig. 7.
It is seen that $t_{opt}$ increases monotonically with increasing $U/V$ from
$t_{opt}=\pi\hbar/2\sqrt{2}V$ at $U/V=0$, see (\ref{t_0_U0}), and approaches
the curve $t_{opt}=\pi\hbar U/4\sqrt{2}V^2$, derived analytically in the
Heisenberg model, for $U>>V$, see (\ref{t_0_t-J}).

It is instructive to examine the dynamics of the inverter upon removing the
external signal from the input dot $A$ at a time $t=T$ without reading the
spin polarization in the output dot $B$. At $t>T$ the electronic subsystem
continues to evolve in accordance with the Schr\"odinger equation
(\ref{Schrodinger_t}). For the sake of simplicity, we restrict our
consideration to the case $U>>V$ (Heisenberg model) since this allows for a
simple analytical solution. For the quantum-mechanically averaged spin
projections at the dots $A$ and $B$ we have
\begin{equation}
S_{zA}(t)=-S_{zB}(t)=S_0\sin\left(\frac{4J(t-T)}{\hbar}+%
\arctan\left(\frac{2J}{\sqrt{H^2+4J^2}}\tan(\omega T/2)\right)\right),
\label{SzA_T}
\end{equation}
where
\begin{equation}
S_0=\frac{H}{H^2+4J^2}\sqrt{\left(H^2\cos^2(\omega T/2)+4J^2\right)%
\sin^2(\omega T/2)}
\label{S_0}
\end{equation}
and $\omega$ is given by Eq. (\ref{omega_2}). We see that $S_{zA}(t)$ and
$S_{zB}(t)$ oscillate between $-S_0$ and $S_0$ having zero average values.
The amplitude of oscillations $S_0$ depends not only on the values of $H$ and
$J$ but on the value of $T$ as well. If $H\ge 2J$ then $S_0$ has its maximum
value 1/2 at $\omega T=\arccos(-4J^2/H^2)+2\pi k$ where $k$ is an integer. It
is interesting that $S_0=0$ at $\omega T=2\pi k$, i.e.,
$S_{zA}(t)=-S_{zB}(t)\equiv 0$, just as in the ground state at $H=0$. In this
case the result of two subsequent perturbations at $t=0$ and $t=T$ is such
that the system at $t>T$ behaves as if it was not perturbed at all. This
effect can be explained in a way that the two perturbations interfere in a
counterphase thus compensating each other.

\vskip 6mm

\centerline{\bf 5. Discussion}

\vskip 2mm

In Sec. 3 we saw that in the presence of input signal at the dot $A$ the
{\it ground state} average of the spin projection
$\langle\hat{S}_{zA}\rangle$ increases monotonically with increasing the
input signal energy $H$ from $\langle\hat{S}_{zA}\rangle=0$ at $H=0$ and
asymptotically approaches the saturation value $\langle\hat{S}_{zA}\rangle=1/2$
at $H>>W$ where $W=Vf(U/V)$ is a characteristic energy which depends on $U$
and $V$ ($W=V$ at $U/V=0$ and $W=J=V^2/U$ at $U/V>>1$, see Fig. 1). In the
case of temporal evolution of electronic subsystem upon application of the
input signal to the dot $A$ one would expect the peak value $S_{zA}(t_0)$ of
the time-dependent spin projection
$S_{zA}(t)=\langle\Psi(t)|\hat{S}_{zA}|\Psi(t)\rangle$ be also governed by
the same parameter, the ratio $H/W$. As follows from analytical expressions
(\ref{SzA(t_0)_U0}) and (\ref{SzA(t_0)_t-J}) obtained for $S_{zA}(t_0)$ in
the limiting cases $U/V<<1$ and $U/V>>1$ respectively, this is so indeed.
However, unlike the ground state average $\langle\hat{S}_{zA}\rangle$, the
dependence of $S_{zA}(t_0)$ on $H/W$ is {\it nonmonotonic}. By contrast,
$S_{zA}(t_0)$ has a maximum $S_{zA}^{max}$ at a {\it finite} value of $H/W$
while tending to zero at $H/W>>1$, see (\ref{SzA(t_0)_U0}),
(\ref{SzA(t_0)_t-J}), Fig. 4, and Fig. 6. Hence there is a fundamental
difference between the physical processes responsible for an increase of the
spin projection in the static and dynamic cases.

We stress that perfect antiferromagnetic spin ordering with maximum absolute
values of spin projections on both dots, $S_{zA}=-S_{zB}=1/2$, can be
realized at the stage of coherent temporal evolution not only in the limit
$U/V\rightarrow\infty$ (this would seem natural by analogy with the static
case when large values of $U/V$ are favorable for antiferromagnetism
\cite{Izyumov}) but also in the limit $U/V\rightarrow 0$. This effect, which
we have called "dynamical antiferromagnetism", arises from the peculiar
nature of electronic transitions under the influence of external
perturbation.

So, in the case of temporal non-dissipative evolution of the inverter, the
maximum permissible (for given $U$ and $V$) value of $S_{zA}$ is achieved
at a {\it finite} value of the input signal energy $H$, i.e.,
$S_{zA}=S_{zA}^{max}$ at $H=H_{opt}$ and $t=t_{opt}$, see Figs. 2 - 4. It is
important that $S_{zA}^{max}\ge 0.45$ at any ratio $U/V$ while
$S_{zA}^{max}\rightarrow 1/2$
at $U/V\rightarrow 0$ or $V/U\rightarrow 0$, see Fig. 5. Since
$S_{zB}(t)=-S_{zA}(t)$ at any time, then if the spin polarization (logical
bit) at the output dot $B$ is measured at a time $t=t_{opt}$, the inversed
input signal, i.e., $S_{zB}=-1/2$ for $H>0$ or $S_{zB}=1/2$ for $H<0$, is
detected with high probability. In other words, the error probability
$P_{err}$, i.e., the probability to measure the wrong output signal
($S_{zB}=1/2$ for $H>0$ or $S_{zB}=-1/2$ for $H<0$) is very low. Indeed, the
upper limit of $P_{err}$ may be estimated as $P_{err}(t)\leq 1-2S_{zA}(t)$,
see Sec. 3. At $t=t_{opt}$ one has $P_{err}(t_{opt})\leq 1-2S_{zA}^{max}$,
i.e., $P_{err}(t_{opt})<<1$ at $U<<V$ or $U>>V$, the maximum value of
$P_{err}(t_{opt})$ being about 0.1 at $U/V=2$.

We recall (see Sec. 3) that in order to reduce $P_{err}$ in ground state
computing, one should increase the input signal energy $H$. In the case of
interest, $P_{err}<<1$, it can be shown from (\ref{SzA-U0}) and
(\ref{SzA-Uinf}) that $H\approx W/\sqrt{P_{err}}$, where $W$ is a
characteristic energy that determines the scale of the energy levels
scheme of the inverter, see Fig. 1. But if so, the value of $H$ should be
also much greater than the thermal energy $k_B T$ since the inequality
$k_B T<<W$ must be fulfilled in order the inverter was in its ground state.
Thus, in fact, one should have
\begin{equation}
k_B T<<W<<H
\label{T<W<H}
\end{equation}
in ground state computing. This condition is obviously hardly realizable
since even the temperature as low as $T=1$ K corresponds to rather high
magnetic field about 1 T, and hence at $T=1$ K two strong inequalities in
(\ref{T<W<H}) can be satisfied only for unrealistic magnetic field about
100 T. Otherwise the error probability $P_{err}$ is very high.

In the case of non-dissipative computing, there is no need for the relaxation
to the new ground state since the perfect antiferromagnetic ordering of
magnetic moments is realized at the stage of unitary evolution when an
appropriate superposition of the inverter eigenstates takes place, see above.
Hence, the input signal energy $H$ should not be anomalously high. For
example, in the limiting case $U>>V$, the optimal value of $H$ is
$H_{opt}=2V^2/U$, so that, e.g., at $V\approx$ 1 meV and $U\approx$ 10 meV
the external magnetic field acting on the input dot can be relatively low,
about 1 T. Such a magnetic field is generated, e.g., by the magnetic moment
of the order of $\mu_B$ in its vicinity. Hence, the input signal may be
thought of as being supplied by the output quantum dot of another gate whose
magnetic moment has been oriented in such way or another at the previous
stage of calculations.

We note that the limiting case $U>>V$ is much more suitable for the physical
implementation than the case $U<<V$. Indeed, at $U<<V$ we have $H_{opt}=2V$,
hence for a realistic value of the input magnetic field about 1 T, the
Coulomb repulsion energy $U$ should not exceed 0.01 meV in order to satisfy
the inequality $U<<V$. Since $U\approx e^2/\varepsilon a$, where $a$ is the
dot size and $\varepsilon$ is the dielectric constant \cite{Openov3}, for
$U\approx$ 0.01 meV one has $a\approx$ 10$^4$ - 10$^5$ nm. Such a relatively
large structure obviously can not be considered as a quantum {\it dot}.

Though the switching of the inverter to the new ground state upon the
influence of the external signal is not necessary in non-dissipative
computing, we stress that our results on unitary evolution of electronic
subsystem of the inverter are obtained under the assumption that at $t\leq 0$
(in the absence of the external signal) the inverter is in its {\it ground
state}. Thus, one faces the problem of preparing such an initial state of the
inverter, and consequently, the problem of temperature reduction, as in the
case of ground state computing. In this respect, it would be interesting to
search for such a type of input signal (e.g., for such a direction of the
external magnetic field, see \cite{Mozyrsky}) whose influence on the input
dot will result in the desired temporal evolution of electronic subsystem
(and consequently, in the realization of the truth table) irrespective of
the initial state of the inverter.

The other facing problem is the reading of the result of computation
(output). While in ground state computing the output can be read at
{\it any } moment after relaxation of the system to its new ground state, in
non-dissipative computing the reading time must be chosen with a great
accuracy as a time when spin projections at the dots have their maximal
values, in order to minimize the error probability. It is interesting to
analyze the relationship between the "optimal" values of $t_{opt}$ and
$H_{opt}$. The dependence of the product $t_{opt}H_{opt}$ on
$U/V$ is shown in Fig. 8. One can see that the value of $t_{opt}H_{opt}$
decreases monotonically with $U/V$ from $\pi\hbar/\sqrt{2}$ at $U/V=0$, see
(\ref{SzA(t_0)_U0}), (\ref{t_0_U0}) down to $\pi\hbar/2\sqrt{2}$ at
$U/V\rightarrow\infty$, see (\ref{SzA(t_0)_t-J}), (\ref{t_0_t-J}). Thus, the
relationship
\begin{equation}
t_{opt}\approx\hbar/H_{opt}
\label{opt}
\end{equation}
holds true for the whole range of $U/V$. In general, this is a consequence of
the Heisenberg's uncertainty principle.

We recall that our purpose is to minimize the switching time $t_{opt}$ and
thereby to maximize the speed of computation. But according to (\ref{opt}),
the decrease in $t_{opt}$ can be achieved only at the sacrifice of increase
in $H_{opt}$. Since the value of the input signal energy is finite, the
relationship (\ref{opt}) imposes a restriction on $t_{opt}$. For realistic
value of external magnetic field $\approx 1$ T one has
$t_{opt}\approx 10^{-11}$ s.

\vskip 6mm

\centerline{\bf 6. Conclusions}

\vskip 2mm

The system consisting of two tunnel-coupled quantum dots, input and output
ones, occupied by two electrons has been considered within the framework of
the Hubbard model with arbitrary values of the parameters, the intra-dot
Coulomb energy $U$ and the inter-dot tunneling energy $V$. After applying the
external signal (the local magnetic field) to the input dot, the magnetic
moments of the dots become oriented in opposite directions, i.e., align
antiferromagnetically. This allows one to use such a system as the quantum
spin gate "inverter" realizing the truth table of logical NOT.

In ground state computing, the inverter switches (relaxes) to the new ground
state under the influence of the input signal. Since the relaxation process
is due to inelastic interactions, the switching time is of the order of
$10^{-9}\div 10^{-6}$ s which is rather long on the atomic scale. Moreover,
the magnetic moments of the dots are far from being saturated, giving rise to
the large probability of reading the wrong result of computation from the
output dot. For these reasons the principle of ground state computing seems
to be ineffective for the spin logical gates.

We have suggested a new approach which overcomes the two above mentioned
drawbacks of ground state computing. Our approach takes advantage of the
coherent non-dissipative evolution of electronic subsystem after applying the
signal to the input dot at $t=0$. We have demonstrated that at {\it any}
moment $t>0$ magnetic moments of the dots arrange in antiferromagnetic order,
i.e., $S_{zA}(t)=-S_{zB}(t)$, thus realizing the truth table of the inverter.
The value of $S_{zA}(t)$ (and hence $S_{zB}(t)$ as well) is a nonmonotonic
function of time, being a periodic function in the limiting cases $U<<V$ and
$U>>V$. At given values of $V$, $U$, and the input signal energy $H$
there always exists a moment $t_0$ which corresponds to the first local
maximum on the curve $S_{zA}(t)$. Besides, there exists an "optimal" value
$H_{opt}$ at which the magnitude of $S_{zA}(t_0)$ reaches its maximum
permissible (for a given ratio of $U/V$) value $S_{zA}^{max}$, thus reducing
the error probability $P_{err}$. Our calculations showed that
$S_{zA}^{max}\geq 0.45$ at any $U$ and $V$, being close to 1/2 at $U<<V$ and
$U>>V$. So, the value of $P_{err}<<1$ is well below than that in ground state
computing. The "optimal" values of $H_{opt}$ and $t_{opt}=t_0(H_{opt})$ are
related by Eq. (\ref{opt}). For reasonable values of $U$, $V$ and realistic
magnetic fields $\sim 1$ T one has $t_{opt}\approx 10^{-11}$ s, several
orders of magnitude less than in ground state computing.

Note that the effect of "dynamical antiferromagnetism" appears to be usable
for further development of more complicated spin gates consisting of a
greater number of quantum dots. However, several problems should be resolved.
First, our results are obtained for a particular case that the initial state
of the inverter is its {\it ground} state. It would be desirable to search
for such a type of the input signal (e.g., for such an orientation of the
external magnetic field with respect to the quantization axis) that the
complete switching of the inverter took place irrespective of its initial
state (the latter may be a superposition of inverter eigenstates). Second,
there is a problem of reading of the result of computation. If the direction
of the magnetic moment on the output dot is measured at a time different from
the "optimal" time $t_{opt}$, then the error probability increases. We note
that the quantum-mechanical average of the spin projection at the output dot
of the inverter has the same sign at any time after the input signal had been
applied to the input dot (this sign determines the value of the logical bit
at the output). Hence, should the electron spin be a classical vector, then
the result of calculation could be determined unambiguously. It would be very
interesting to search for such a measurement procedure that enabled us to
draw a conclusion about the {\it average} direction of the magnetic moment
over some period of time. Solving of these problems appears to be the target
of future research.

\vskip 6mm

\centerline{\bf Acknowledgments}

\vskip 2mm

This work was supported in part by the Russian Foundation for Fundamental
Research under Grant No 96-02-18918 and by the Russian State Program
"Advanced Technologies and Devices in Micro- and Nanoelectronics" under
Grant No 02.04.329.89.5.3. We are grateful to S.Bandyopadhyay for sending us
a preprint of the work \cite{Bandyo2} prior to publication and to
I.A.Semenihin for the help at the early stage of this work. We would like
to thank A.V.Krasheninnikov, S.N.Molotkov, and S.S.Nazin
for fruitful discussions.

\newpage
\centerline{\bf Figure captions}
\vskip 2mm

Fig.1. The function $f(x)$ entering into the expression for the ground state
average of the spin projection
$\langle \hat{S}_{zA}\rangle=H/2\sqrt{H^2+4V^2f^2(U/V)}$
at the quantum dot $A$.

Fig.2. Temporal evolution of the quantum-mechanically averaged spin
projection $S_{zA}(t)=\langle \Psi (t)|\hat{S}_{zA}|\Psi (t)\rangle$ at the
quantum dot $A$ upon applying the local external magnetic field $H_A$ to the
dot $A$. The intradot Coulomb repulsion energy $U=0$ at both dots $A$ and $B$
(the weak coupling limit of the Hubbard model). $H=V$ (dashed line), $H=2V$
(solid line), $H=6V$ (dotted line). Here $H=g\mu_B H_A/2$ is the input signal
energy, $V$ is the energy of electron tunneling between the quantum dots, the
time $t$ is measured in units of $\hbar/V$.

Fig.3. The same as in Fig.2, for a) $U/V=2$, $H/V=2$; b) $U/V=1$, $H/V=4$;
c) $U/V=1.5$, $H/V=1.155$.

Fig.4. The peak value of $S_{zA}(t_0)$ versus $H/V$ at different $U/V$.
The switching time $t_0$ is defined as a time of the first peak at the curve
$S_{zA}(t)$ at given values of $U/V$ and $H/V$.

Fig.5. The dependence of $S_{zA}^{max}$ on $U/V$. Here $S_{zA}^{max}$ is the
maximum value of $S_{zA}(t_0)$ on the $S_{zA}(t_0)$ versus $H/V$ curve at a
given $U/V$.

Fig.6. The "optimal" value of normalized input signal energy $H_{opt}/V$
versus $U/V$. Here $H_{opt}$ is the value of $H$ at which the peak value
$S_{zA}(t_0)$ reaches its maximum $S_{zA}^{max}$ at a given $U/V$, see
Fig.4. Solid line is the result of numerical calculations. Dashed line is
the curve $H_{opt}/V=2V/U$ obtained in the Heisenberg model, i.e., for
$U>>V$.

Fig.7. The "optimal" switching time $t_{opt}$ measured in units of $\hbar/V$
as a function of $U/V$. Here $t_{opt}$ is the time of the first peak on the
curve $S_{zA}(t)$ at $H=H_{opt}$, i.e., $t_{opt}(U/V)=t_0(U/V,H_{opt}/V)$
corresponds to $S_{zA}^{max}$, the maximum permissible value of $S_{zA}(t_0)$
at a given $U/V$. Solid line is the result of numerical calculations. Dashed
line is the curve $t_{opt}V/\hbar=\pi U/4\sqrt{2}V$ obtained in the
Heisenberg model, i.e., for $U>>V$.

Fig.8. The product of "optimal" switching time $t_{opt}$ measured in units
of $\hbar/V$ by the "optimal" value of normalized input signal energy
$H_{opt}/V$ versus $U/V$.

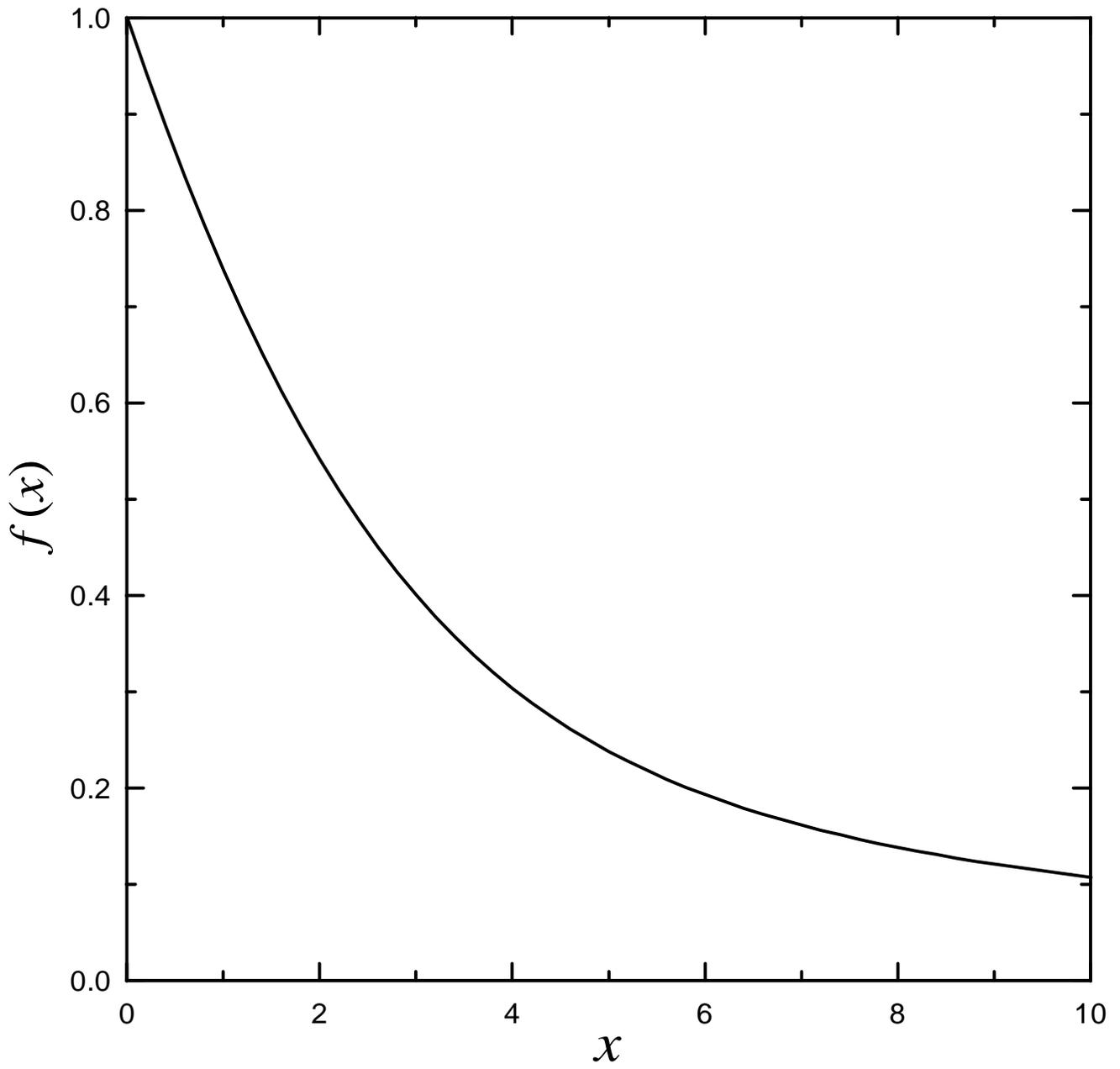
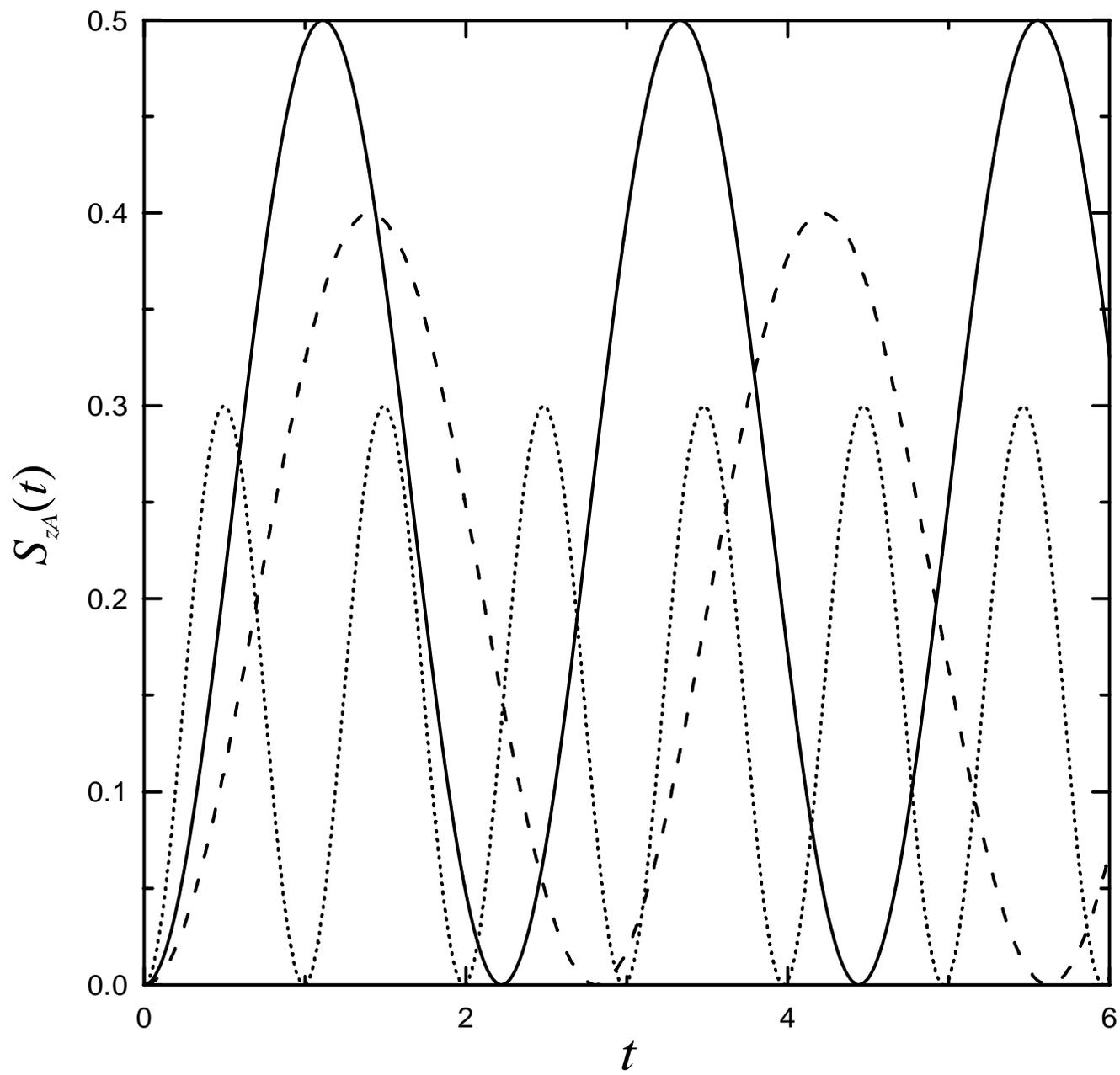
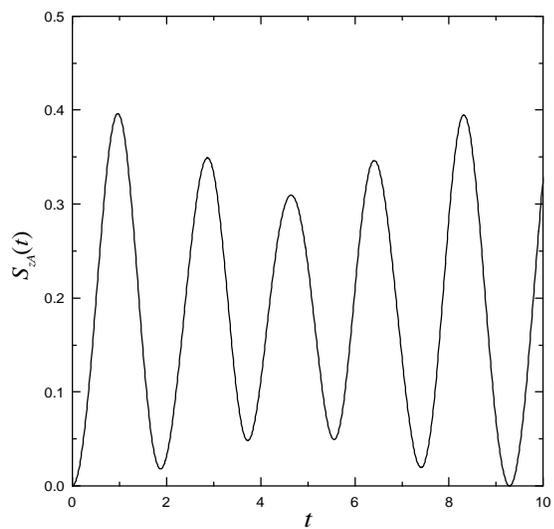
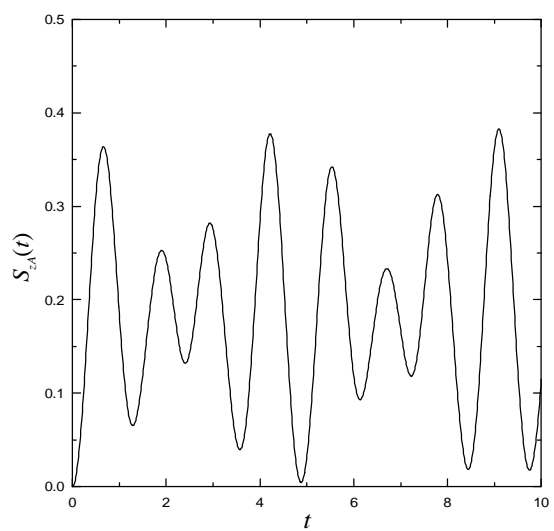
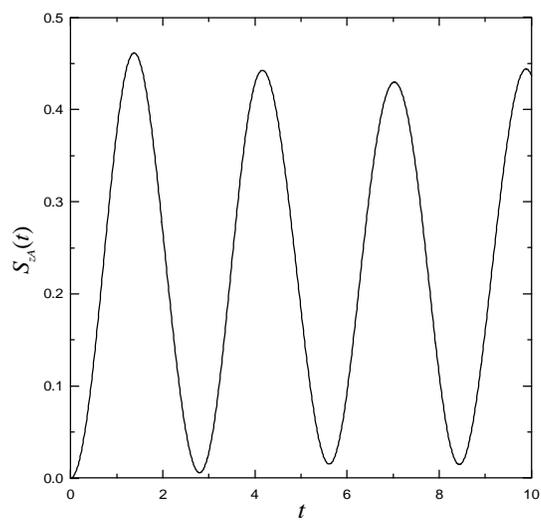
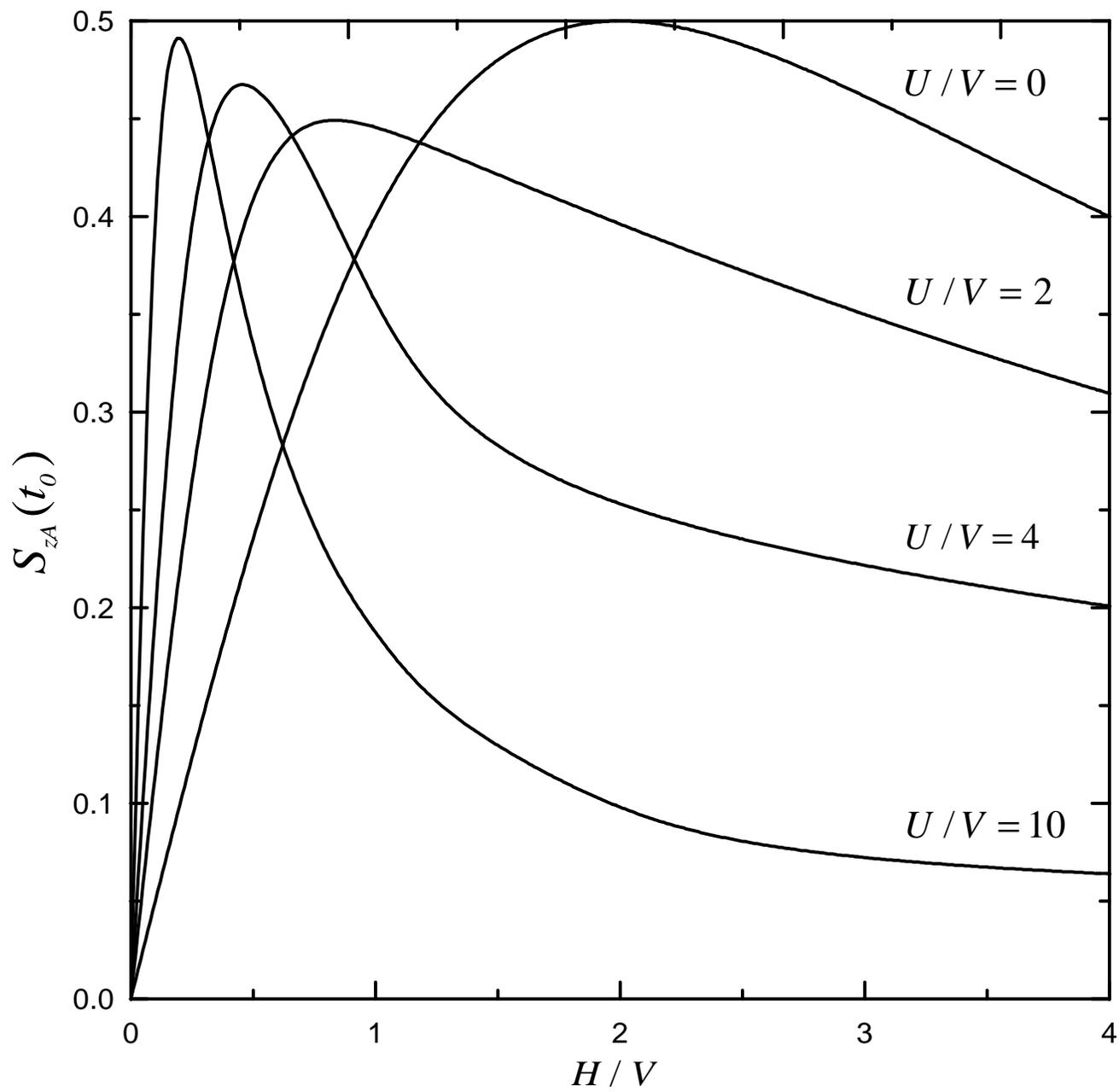
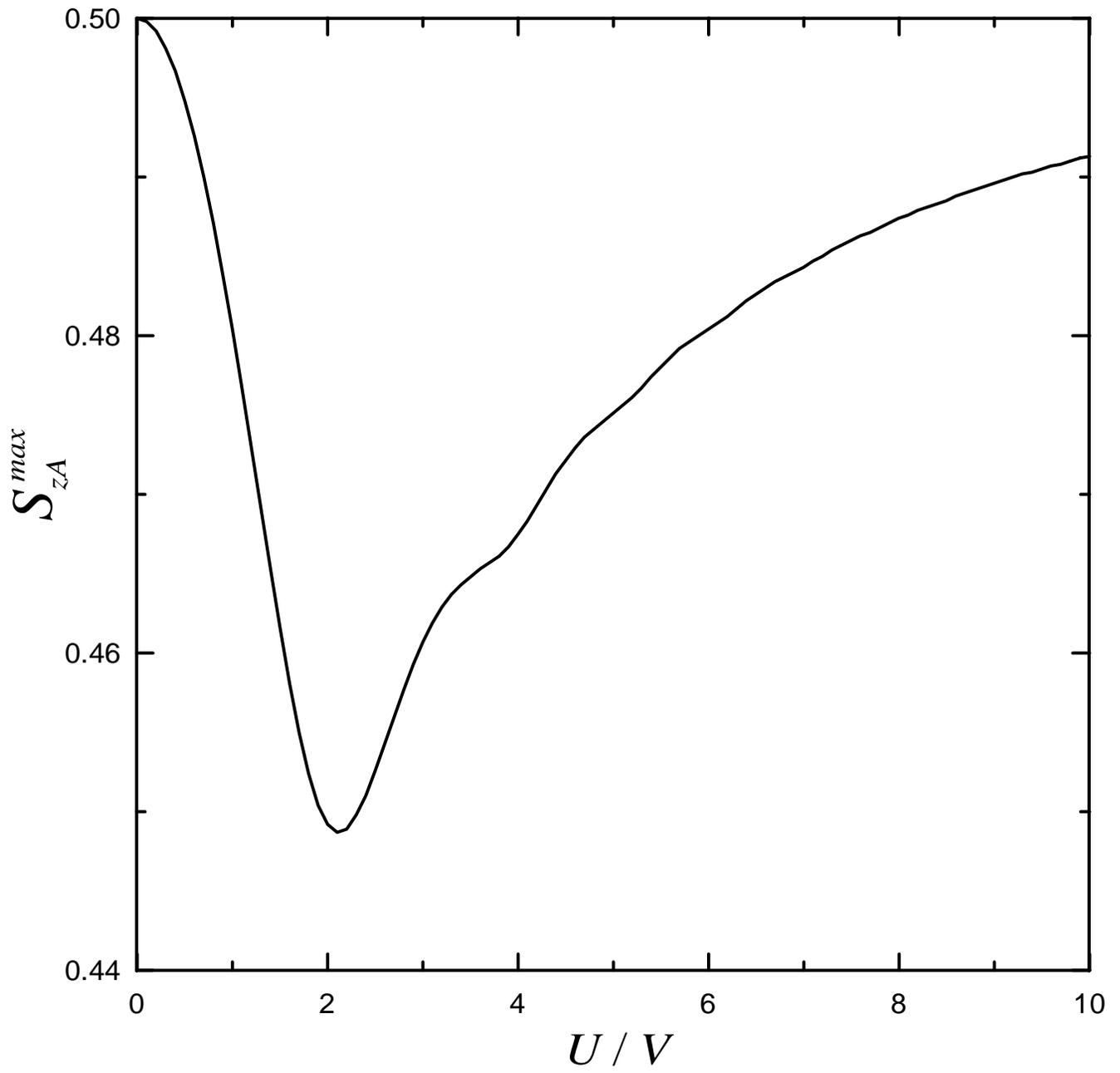
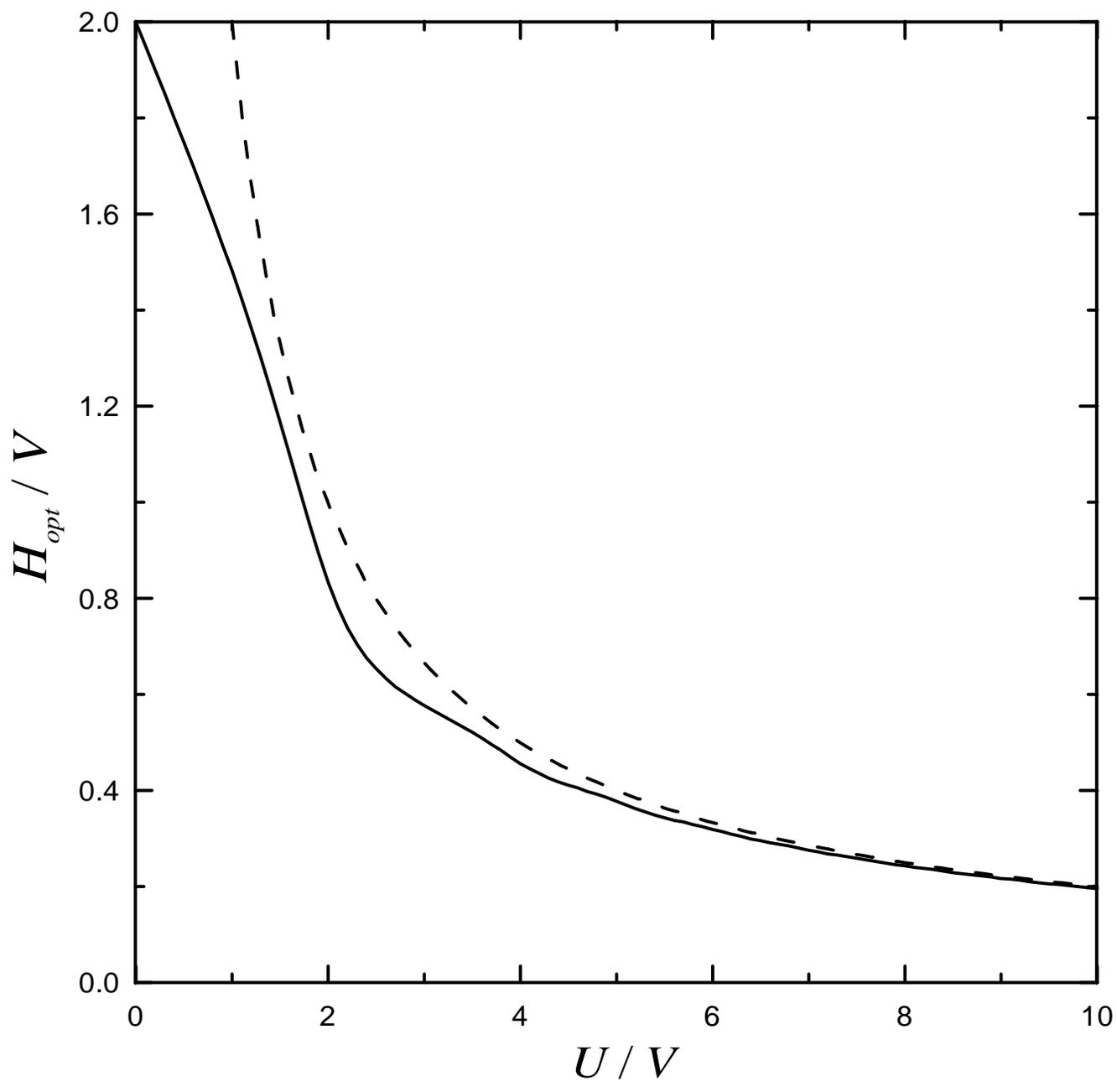
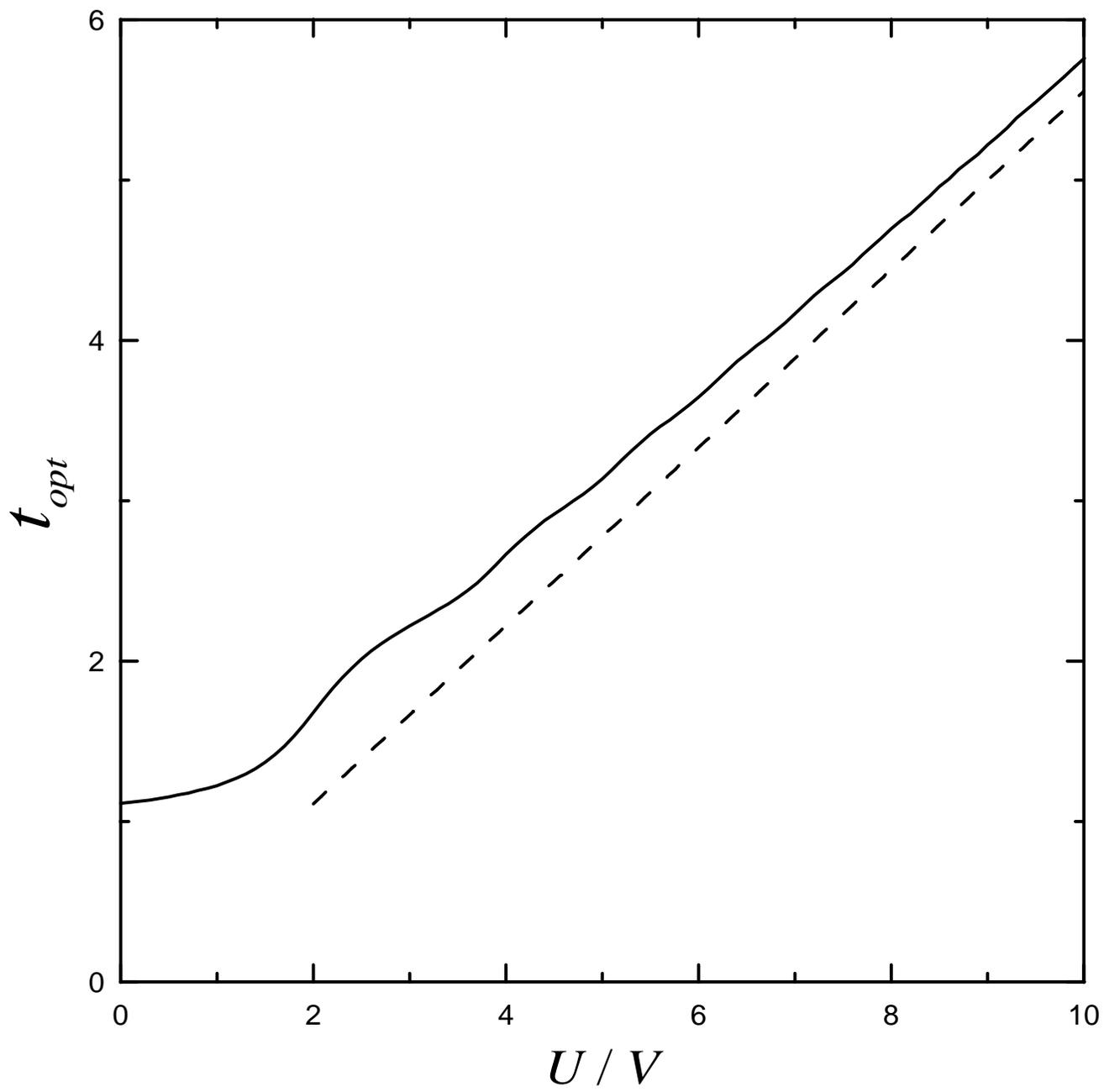
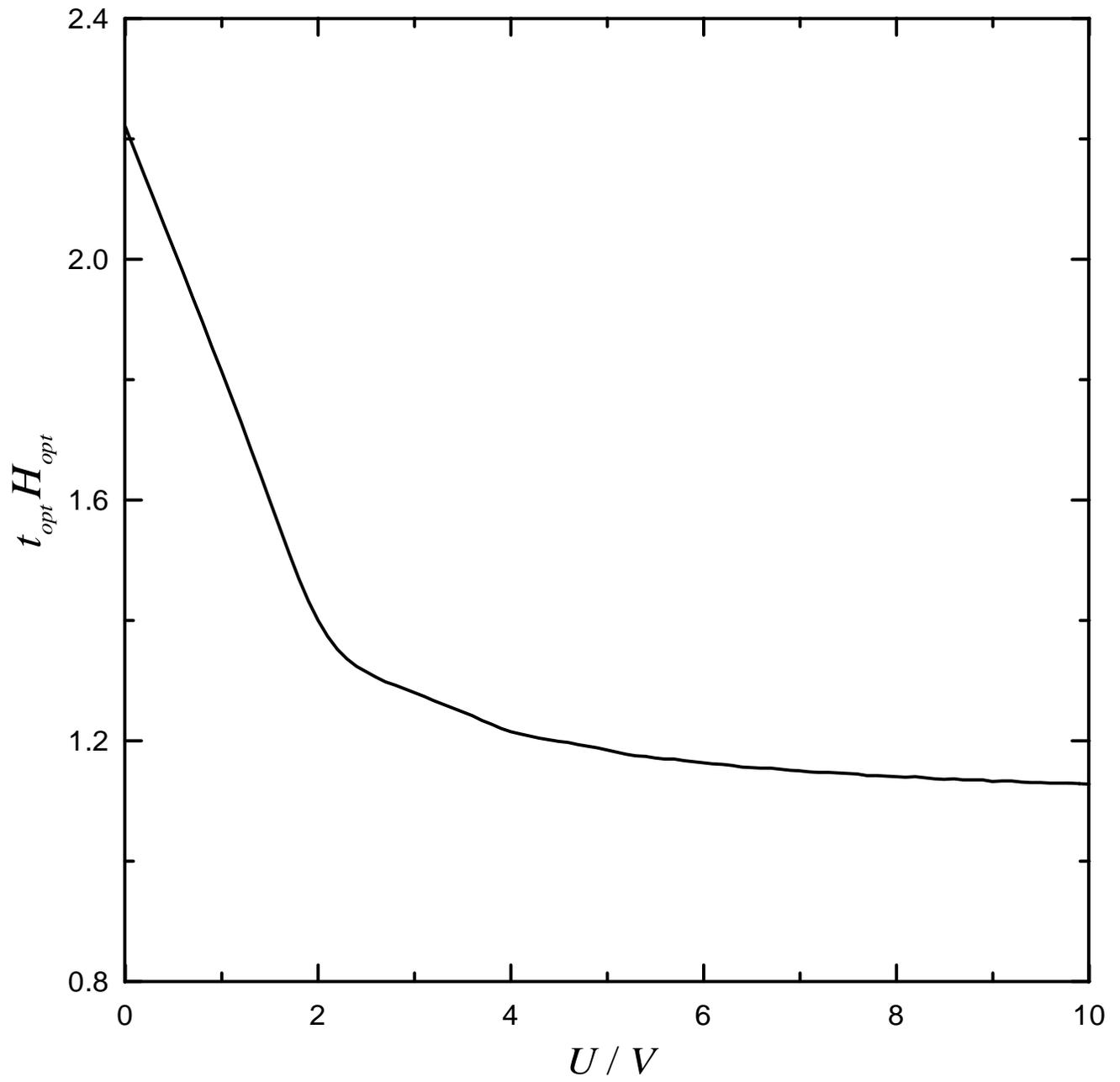
\end{document}